\documentclass[conference]{IEEEtran}
\usepackage{cite}
\usepackage{amsmath,amssymb,amsfonts}
\usepackage{algorithmic}
\usepackage{graphicx}
\usepackage{textcomp}
\usepackage{xcolor}
\usepackage{fancyhdr}
\usepackage[hyphens]{url}
\usepackage{booktabs}
\usepackage{multirow}
\usepackage{amsthm}

\usepackage[symbol]{footmisc}

\newlength{\Oldarrayrulewidth}
\newcommand{\Cline}[2]{%
  \noalign{\global\setlength{\Oldarrayrulewidth}{\arrayrulewidth}}
  \noalign{\global\setlength{\arrayrulewidth}{#1}}\cline{#2}%
  \noalign{\global\setlength{\arrayrulewidth}{\Oldarrayrulewidth}}}

\usepackage[capitalise,noabbrev]{cleveref}
\usepackage{subfig}

\def\BibTeX{{\rm B\kern-.05em{\sc i\kern-.025em b}\kern-.08em
    T\kern-.1667em\lower.7ex\hbox{E}\kern-.125emX}}

\title{
    \vspace{-2mm}
    First-Generation Inference Accelerator Deployment at Facebook
    \vspace{-6mm}
}

\author{
\IEEEauthorblockN{
\small
Michael Anderson,
Benny Chen,
Stephen Chen,
Summer Deng,
Jordan Fix,
Michael Gschwind,
Aravind Kalaiah,
Changkyu Kim,
Jaewon Lee,
\\
Jason Liang,
Haixin Liu,
Yinghai Lu,
Jack Montgomery,
Arun Moorthy,
Satish Nadathur,
Sam Naghshineh,
Avinash Nayak,
Jongsoo Park,
\\
Chris Petersen,
Martin Schatz,
Narayanan Sundaram,
Bangsheng Tang,
Peter Tang,
Amy Yang,
Jiecao Yu,
Hector Yuen,
Ying Zhang,
\\
Aravind Anbudurai,
Vandana Balan,
Harsha Bojja,
Joe Boyd,
Matthew Breitbach,
Claudio Caldato,
Anna Calvo,
Garret Catron,
Sneh Chandwani,
\\
Panos Christeas,
Brad Cottel,
Brian Coutinho,
Arun Dalli,
Abhishek Dhanotia,
Oniel Duncan,
Roman Dzhabarov,
Simon Elmir,
Chunli Fu,
\\
Wenyin Fu,
Michael Fulthorp,
Adi Gangidi,
Nick Gibson,
Sean Gordon,
Beatriz Padilla Hernandez,
Daniel Ho,
Yu-Cheng Huang,
Olof Johansson,
\\
Shishir Juluri,
Shobhit Kanaujia,
Manali Kesarkar,
Jonathan Killinger,
Ben Kim,
Rohan Kulkarni,
Meghan Lele,
Huayu Li,
Huamin Li,
\\
Yueming Li,
Cynthia Liu,
Jerry Liu,
Bert Maher,
Chandra Mallipedi,
Seema Mangla,
Kiran Kumar Matam,
Jubin Mehta,
Shobhit Mehta,
\\
Christopher Mitchell,
Bharath Muthiah,
Nitin Nagarkatte,
Ashwin Narasimha,
Bernard Nguyen,
Thiara Ortiz,
Soumya Padmanabha,
Deng Pan,
\\
Ashwin Poojary,
Ye (Charlotte) Qi,
Olivier Raginel,
Dwarak Rajagopal,
Tristan Rice,
Craig Ross,
Nadav Rotem,
Scott Russ,
Kushal Shah,
\\
Baohua Shan,
Hao Shen,
Pavan Shetty,
Krish Skandakumaran,
Kutta Srinivasan,
Roshan Sumbaly,
Michael Tauberg,
Mor Tzur,
Sidharth Verma,
\\
Hao Wang,
Man Wang,
Ben Wei,
Alex Xia,
Chenyu Xu,
Martin Yang,
Kai Zhang,
Ruoxi Zhang,
Ming Zhao,
Whitney Zhao,
Rui Zhu,
\\
Ajit Mathews,
Lin Qiao,
Misha Smelyanskiy,
Bill Jia,
Vijay Rao
}
\vspace{1mm}
\IEEEauthorblockA{Facebook, Inc.}
\vspace{-7mm}
}

\begin{document}
\maketitle
\pagestyle{plain}


\begin{abstract}

In this paper, we provide a deep dive into the deployment of inference accelerators at Facebook. Many of our ML workloads have unique characteristics,  such as sparse memory accesses, large model sizes, as well as high compute, memory and network bandwidth requirements. We co-designed a high-performance, energy-efficient inference accelerator platform based on these requirements. We describe the inference accelerator platform ecosystem we developed and deployed at Facebook: both hardware, through Open Compute Platform (OCP), and software framework and tooling, through Pytorch/Caffe2/Glow. A characteristic of this ecosystem from the start is its openness to enable a variety of AI accelerators from different vendors. This platform, with six low-power accelerator cards alongside a single-socket host CPU, allows us to serve   models of high complexity that cannot be easily or efficiently run on CPUs. We describe various performance optimizations, at both platform and accelerator level, which enables this platform to serve production traffic at Facebook. We also share deployment challenges, lessons learned during performance optimization, as well as provide guidance for future inference hardware co-design.

\end{abstract}
\section{Introduction}
\label{section:intro}

\begin{figure}[t]
    \centering
    \includegraphics[width=0.85\columnwidth]{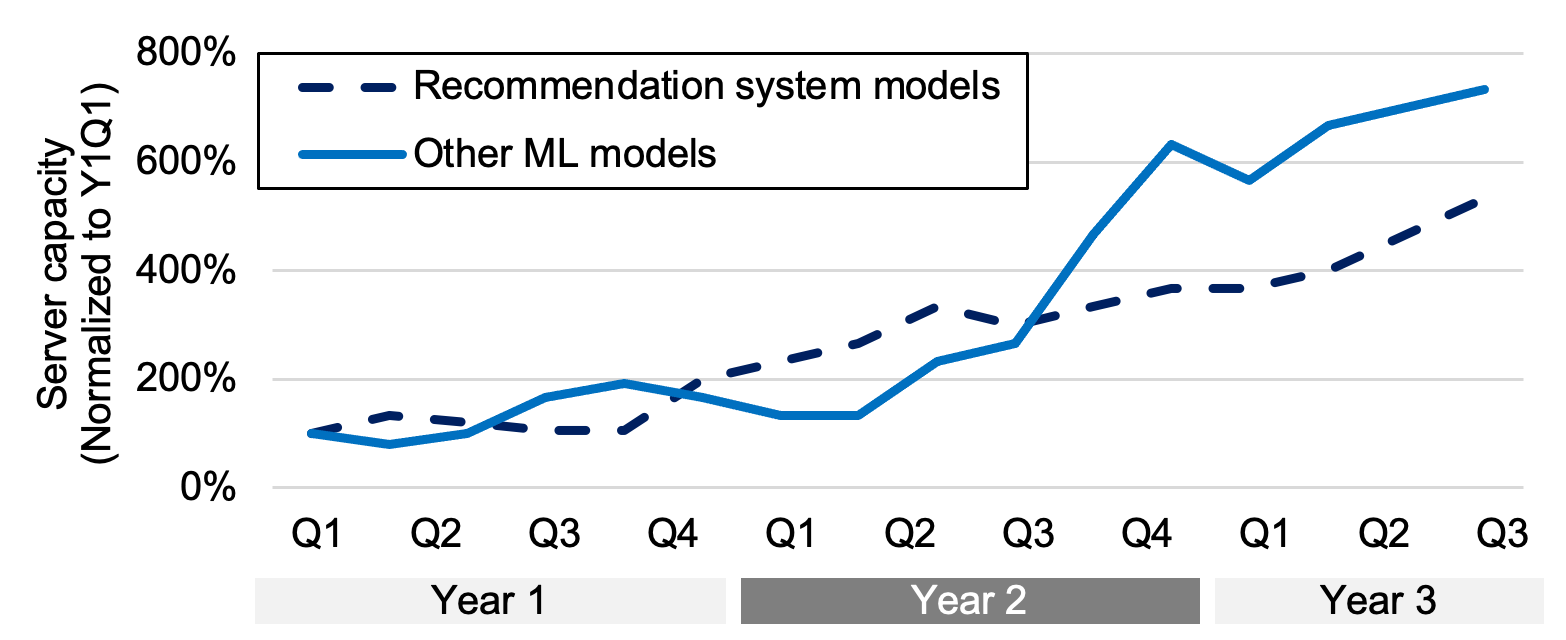}
\vspace{-2mm}
\caption{Growth in number of servers for inference of (a) recommendation system models and (b) other ML models such as computer vision and text.}
\vspace{-6mm}
\label{fig:growth}
\end{figure}

Machine learning models at Facebook are placing increasingly heavy demands on our data centers, which have been primarily based on CPUs \cite{park2018deep}. There has been 5$\sim$7$\times$ growth in the number of servers used for ML inference over a span of 2 years as shown in \cref{fig:growth}; and this is projected to increase at similar rates in the future. Moreover, our models are becoming more complex in terms of FLOPs and model size. For instance, Natural Language Processing (NLP) models have been developed which are over 300$\times$ larger than those we are deploying currently \cite{brown2020language}. Inference of such complex models on CPUs may stretch latency budgets. Consequently, Facebook has deployed inference accelerators to meet both latency and throughput needs of our models.

For the inference accelerator program at Facebook,  both  hardware  and  software  platforms could  be  developed  from  scratch  and  co-designed  for  our needs. First, the system needed to be capable of achieving a high perf/watt ratio, targeting a range of ML applications. Second, the system needed to be able to support large recommendation system models \cite{naumov2019deep, arch_implications_facebook_dnn_rec, instagram_model} (e.g. many tens of GB even after model optimization) without sacrificing performance. These models require large memory and also high bandwidth for fast lookups. Finally, the system needed to be able to adapt to a fast-evolving landscape of ML frameworks (e.g. Caffe2 to PyTorch \cite{caffe2pytorch}), and inference accelerator offerings from different vendors.

In order to address these challenges, we co-designed both hardware and software aspects of our system. We started with a building block of a low-power accelerator card with high peak TOPS/W. The card offers a peak performance, deriving from frequencies which range depending on the workload, of 30-45 TOPS (int8) or 4-6 TFLOPS (FP16), has 16 GB of LPDDR memory, and consumes only 13W.  In order to increase memory capacity, memory bandwidth, and compute capability of the system, we connect six cards through a PCIe switch. These six cards can either handle independent requests, or alternatively function as a combined large system with a peak performance of 180-270 TOPS (int8) or 24-36 TFLOPS (FP16), with 96 GB of LPDDR memory, running at 91W including the PCIe switch. The combined system is sufficiently sized to handle large recommendation system workloads, provided we use a custom multi-card partitioning strategy specific to recommendation systems. Overall the hardware peak efficiency is 2.0-3.0 TOPS/W. 

We are building the Facebook inference accelerator ecosystem with a focus on openness from the start \cite{yosemite_article}.  Hardware specifications are made available through OCP, with a focus on enabling a variety of AI accelerators. Frameworks (PyTorch, Caffe2) and AI accelerator tooling (Glow, onnxifi\cite{rotem2019glow}) are available as open source software and already support a number of different execution platforms and accelerators. Finally, we have open sourced models for recommendation system and content understanding, such as DLRM\cite{naumov2019deep}, PyText \cite{pytext}, and Classy Vision \cite{adcock2019classy}, including the necessary accelerator support. The focus on openness benefits both Facebook and the AI community by creating a common ecosystem in which to innovate, ranging from hardware accelerators to AI compilation to model definition, and in turn, advancing the state of the art in AI acceleration to create higher quality, larger models with the potential to deliver qualitative improvements in outcomes.

To co-design the software for high performance, we performed various optimizations at the \emph{model-level} (quantization and data-type changes for compute, co-design to avoid expensive ops on the inference cards, etc.); \emph{card-level} (graph level optimizations such as model partitioning among cards, resource allocation, parallelization across cores and placement of ops to cores, batching to improve memory re-use, etc);  \emph{op-level} (optimization to improve the performance of individual operators); as well as \emph{system level} (changes in the serving stack to minimize communication to the accelerator, improving host CPU utilization and PCIe traffic optimizations). 

As a result, we are able to serve complex models on the accelerator while still meeting our application latency requirements. The goal of using larger and more complex models is to increase model accuracy. For recommendation systems, we can serve a model which is 5$\times$ more complex in terms of GFLOPS and has 2$\times$ more parameters than current models. The accelerator system also handles computer vision models efficiently: for a complex RegNetY model \cite{radosavovic2020designing}, which is 15$\times$ more complex in both GFLOPS and parameters than current  models, also meets the latency requirement on accelerators. For NLP the complex XLM-R variant has 6$\times$ more compute in GFLOPS than current models and also meets latency requirements on the accelerator. These particular complex models are well suited for the accelerator and cannot be easily or efficiently run on CPU. However CPUs are still valuable for many other inference tasks and continue to improve with the addition of low precision and AI instruction support.

The paper is organized as follows. 
We first describe the inference workloads at Facebook (\cref{section:workloads}).
We then describe the hardware system that we developed at both the system and individual card level (\cref{section:hardware}), followed by the SW design (\cref{section:software}).
We then look at performance considerations, including quantization and numerics (\cref{section:numerics})
and optimizations at the model, card and system level (\cref{section:optimizations}).
We show the performance results of running complex models on the optimized system (\cref{section:results}), 
and share insights into how inference hardware design should be adapted for future AI workloads (\cref{section:directions}).
Finally, we describe related studies (\cref{section:relatedwork}) and conclude the paper (\cref{section:conclusion}).

\section{Inference Workloads at Facebook}
\label{section:workloads}

\setlength{\aboverulesep}{0pt}
\setlength{\belowrulesep}{0pt}
\setlength\heavyrulewidth{0.25ex}

\begin{table*}[ht!]
\centering
\scriptsize
\caption{Model Characteristics.}
\vspace{-2.5mm}
\renewcommand*{\arraystretch}{0.93}
\renewcommand*{\tabcolsep}{3pt}
\begin{tabular}{c|c|c|c|c|c|c}
\toprule
\textbf{Category} & \textbf{Model name} & \textbf{\begin{tabular}[c]{@{}c@{}}Model Size (MParams)\end{tabular}} & \textbf{\begin{tabular}[c]{@{}c@{}}FLOPs\\ (GFLOPs per batch)\end{tabular}}  & \textbf{\begin{tabular}[c]{@{}c@{}}Typical Batch\\ Size\end{tabular}} & \textbf{\begin{tabular}[c]{@{}c@{}}Arith. Intensity\\ (Weights+Activations)\end{tabular}}  & \textbf{\begin{tabular}[c]{@{}c@{}}Latency\\ Constraints (ms)\end{tabular}} \\ \toprule
\multirow{2}{*}{Recommendation} & Less complex model & 70,000  & 0.02 & 32-64 & 90  & 100 (per 150-180 items) \\ \cline{2-7}
 & More complex model & $>$100,000  & 0.1 & 32-64 & 80  & 100 (per 150-180 items)\\ \hline
\multirow{3}{*}{Computer Vision} & ResNeXt101-32x4-48 & 44 & 15.6 & 1 image & 355  & $\sim$1000 \\ \cline{2-7}
 & RegNetY & 700 & 256 & 1 image & 395 & $\sim$1000 \\ \cline{2-7}
 & FBNetV3 based model & 28.6 & 72 & 1 image & 1946 & $\sim$300 \\ \hline
Video Understanding & ResNeXt3D based & 58 & 3.4 & 4 frames & 362 & $\sim$350 \\ \hline
\begin{tabular}[c]{@{}c@{}}Natural Language\\Processing \end{tabular} & XLM-R & \begin{tabular}[c]{@{}c@{}}558\end{tabular} & \begin{tabular}[c]{@{}c@{}}20\\ (32 tokens)\end{tabular} & \begin{tabular}[c]{@{}c@{}}20-70 tokens\\ (1 sentence)\end{tabular} & \begin{tabular}[c]{@{}c@{}}\#tokens\\ (20-70)\end{tabular}  & $\sim$200 \\
\bottomrule
\end{tabular}
\vspace{-5.5mm}
\label{table:models}
\end{table*}

There are four classes of workloads we are running: (1) personalized recommendation models based on prior user interactions; (2) computer vision models for image classification and object detection; (3) language understanding models that feed into applications such as integrity use-cases, and (4) video understanding models. The last three classes are Content Understanding models used in applications such as finding objectionable content. From a computational perspective, the overall characteristics of these models that we analyze in this paper are listed in Table 1. While the broad classes of workloads have not changed since they were outlined in prior work~\cite{park2018deep}, specifics of the models have changed. Latency requirements for recommendation systems have not changed significantly as ~\cite{park2018deep} was reporting per-batch requirements. Overall embedding table sizes have grown from tens of billions to hundreds of billions of params in the three-year span. Latencies for the NLP model has since been relaxed due to nearline processing, and the NLP model architecture has shifted from GRU/LSTM to Transformer. We go into further details on each of these models below.

\subsection{Recommendation System Workloads}
\noindent\textbf{Use cases}:
Recommendation models are a common type of model used in various applications such as ranking and search. The basic structure of recommendation models used at Facebook is a Deep Learning Recommendation Model as described in~\cite{naumov2019deep, arch_implications_facebook_dnn_rec, instagram_model}.

\begin{figure}[t]
    \centering
    \includegraphics[width=1.0\columnwidth]{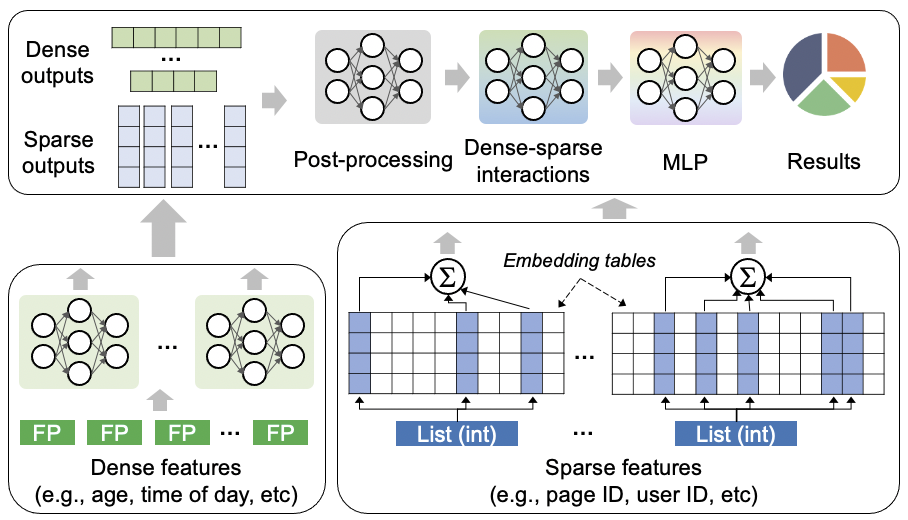}
    \vspace{-6mm}
    \caption{Structure of Deep Learning Recommendation Model (DLRM).}
    \vspace{-6mm}
    \label{fig:dlrm}
\end{figure}

The structure of the model is illustrated in \cref{fig:dlrm}. The model takes as input a set of sparse (categorical) and dense (continuous) features. The sparse features are mapped to a dense representation in an abstract space using embedding table lookups, known as Sparse Lengths Sum (SLS) in Caffe2 (EmbeddingBag in PyTorch), followed by a pooling operation. The dense features are then processed through Multi-Layer Perceptron (MLP) or Fully Connected (FC) layers, and then interactions are computed with the dense features~\cite{factorization_machines}.

\vspace{1mm}
\noindent\textbf{Characteristics and impact on platform design}:

The primary characteristic of these recommendation systems is the extremely large embedding tables which contain over 100B parameters, and must be available for fast lookups. This memory capacity requirement is significantly larger than other applications. Many of the embedding tables are quantized in order to fit more parameters on the accelerator cards. By contrast, the dense compute layers contain relatively small number of weights, in the tens of millions of parameters, and also relatively low in arithmetic intensity of 80-90 ops per byte. These compute layers therefore would benefit greatly from weights storage in on-chip memory. The interaction layers combine the results of the dense compute layers with the results of the embedding lookups, meaning that the results of the embedding lookups must be gathered into one place, and therefore cannot be entirely processed in a distributed manner. 
Finally, at a system level, we ensure not to be bottlenecked by the network interface (NIC) since the input features to these models can be fairly large.

\vspace{-1mm}
\subsection{CV Workloads}
\vspace{-1mm}
\noindent\textbf{Use cases}: Computer Vision models have used deep learning models for quite some time to perform tasks such as image classification and object detection. 

Image classification involves generating a class for each image. Currently, image classification models have relied on ResNeXt type models to produce state of the art accuracy~\cite{mahajan2018exploring,xie2017aggregated}. However, the most recent models have used much larger and computationally complex RegNet models~\cite{radosavovic2020designing}. In particular, we use a ResNeXt-101 model and a RegNetY model in this work. The RegNetY model has more than an order of magnitude more parameters and FLOPs than the ResNext-101 model, and hence cannot be served within the latency constraint without using an accelerator.

Object detection models involve identifying specific regions that contain objects of interest, for example to detect text in images~\cite{rosetta}. Here we use a recent architecture produced by Neural Architecture Search, called FBnetv3~\cite{dai2020fbnetv3} as the backbone. This uses a mix of channelwise and regular convolutional layers. There are some additional ops that generate region proposals followed by a classification net at the end.

\vspace{1mm}
\noindent\textbf{Characteristics and impact on accelerator design}: These models are fairly heavy in terms of compute. They consist primarily of convolutions which have high arithmetic intensity, but can also have groupwise/channelwise convolutions which have reduced arithmetic intensity. The RegNetY model is much larger than the others. At $\sim$1 GB even in quantized form, RegNetY is unlikely to fit into on-chip memory, and is therefore more sensitive to memory bandwidth performance.   

\vspace{-1mm}
\subsection{NLP Workloads}
\vspace{-1mm}
\noindent\textbf{Use cases}:
NLP models have changed significantly in recent years with transformer based models~\cite{radford2018improving} showing state-of-the-art accuracy. We use a model based on XLM-R~\cite{conneau2020unsupervised}, a cross-language Transformer-based language model which is trained on multiple terabytes of data in over one hundred languages. This model has been used for a variety of tasks, including Cross-Lingual Classification, Named Entity Recognition and Cross-Lingual Language Understanding~\cite{conneau2020unsupervised}. Transformer-based models are composed primarily of attention based mechanisms; the model we use has 24 layers of XLM-R with a total of 558 million parameters.

\vspace{1mm}
\noindent\textbf{Characteristics and impact on accelerator design}:
These models have a large portion of runtime devoted to matrix-multiply compute, and are thus particularly well suited for AI accelerators. However, the matrix-multiply layers involved have low arithmetic intensity (equal to the batch size times the average number of input tokens), which in our use cases is somewhere between 20 and 70. At this range, the layers are limited by memory bandwidth rather than compute.

The sequence length, or the number of tokens per sentence, can vary depending on the application between one to several hundred, though smaller lengths are more common. This means that the shapes of the inputs to the operator will vary. For accelerators that require static shapes for compilation, one needs to pad the inputs up to one of a small set of upper-bound sequence lengths (e.g. 32, 64, 96, ... 256), compile a separate network for each possible size, and be able to quickly switch between networks at runtime. 

Transformer models have been growing larger and more complex; the XLM-R model we use already is large enough at $\sim$1 GB in FP16 that it is unlikely to fit in on-chip memory. NLP models have been developed with tens or even hundreds of billions of parameters~\cite{sharir2020cost,brown2020language}. What stands in the way of deployment is computational cost, in terms of aggregate capacity and latency. At present, high-performance inference accelerators offer the optimal path to deployment with acceptable latency and cost. Even with accelerators, to handle these much larger models, we will need to distribute the model among multiple accelerator cards in a single machine or even perhaps across multiple machines. 

\vspace{-1mm}
\subsection{Video Workloads}
\vspace{-1mm}
\noindent\textbf{Use cases}:
Video workloads are characterized by their unique requirements for decode and real-time inference on live videos. Processing every frame of a video with this requirement is resource intensive. Instead, we sample 3D clips of the video at regular intervals and process it at a reduced spacial resolution using a model based on ResNext3D~\cite{Tran_2019_ICCV} and Octave Convolution~\cite{Chen_2019_ICCV}. Furthermore, we run additional models for speech and character recognition on the audio and visual frames. The output from these individual trunks are fused to create a multi-modal model~\cite{Wang_2020_CVPR} that generates embeddings that are used for classification and for downstream AI tasks. 

\noindent\textbf{Characteristics and impact on accelerator design}:
The inference time on a second of video cannot take longer than a second. So there is a latency-driven tradeoff between the complexity of the model, the number of models and the video sampling parameters. 

The video trunk is dominated by 3D convolutions (regular $1\times 1\times 1$ cross-channel convolution and $3\times 3\times 3$ depth-wise convolution) as well as bandwidth bound operations such as pooling and batch normalization. Since the output activation size tends to be high, it is important for the accelerator to be able to fuse bandwidth-bound ops with compute ops and to avoid resource intensive data transformations. Decoding of compressed video frames involves a significant portion of the compute and is typically done on the CPU and can become a bottleneck. This makes for a tight coupling between the network I/O, video decode and running several models under a tight latency bound. So the accelerated video inference ecosystem needs to be heterogeneous and support streaming distributed inference across several models. 
\section{Hardware Design}
\label{section:hardware}

The unique requirements from our inference workloads led us to co-design the system to maximize efficiency. As such, we designed a system in-house \cite{yosemite_article}, comprising of a host CPU with multiple inference accelerator cards. The specifications for this system and its components were released to the Open Compute Platform \cite{yosemite_ocp}. We describe the platform as well as the accelerator card design below.

\subsection{System Design}
\begin{figure}[t]
    \centering
    \includegraphics[width=0.8\columnwidth]{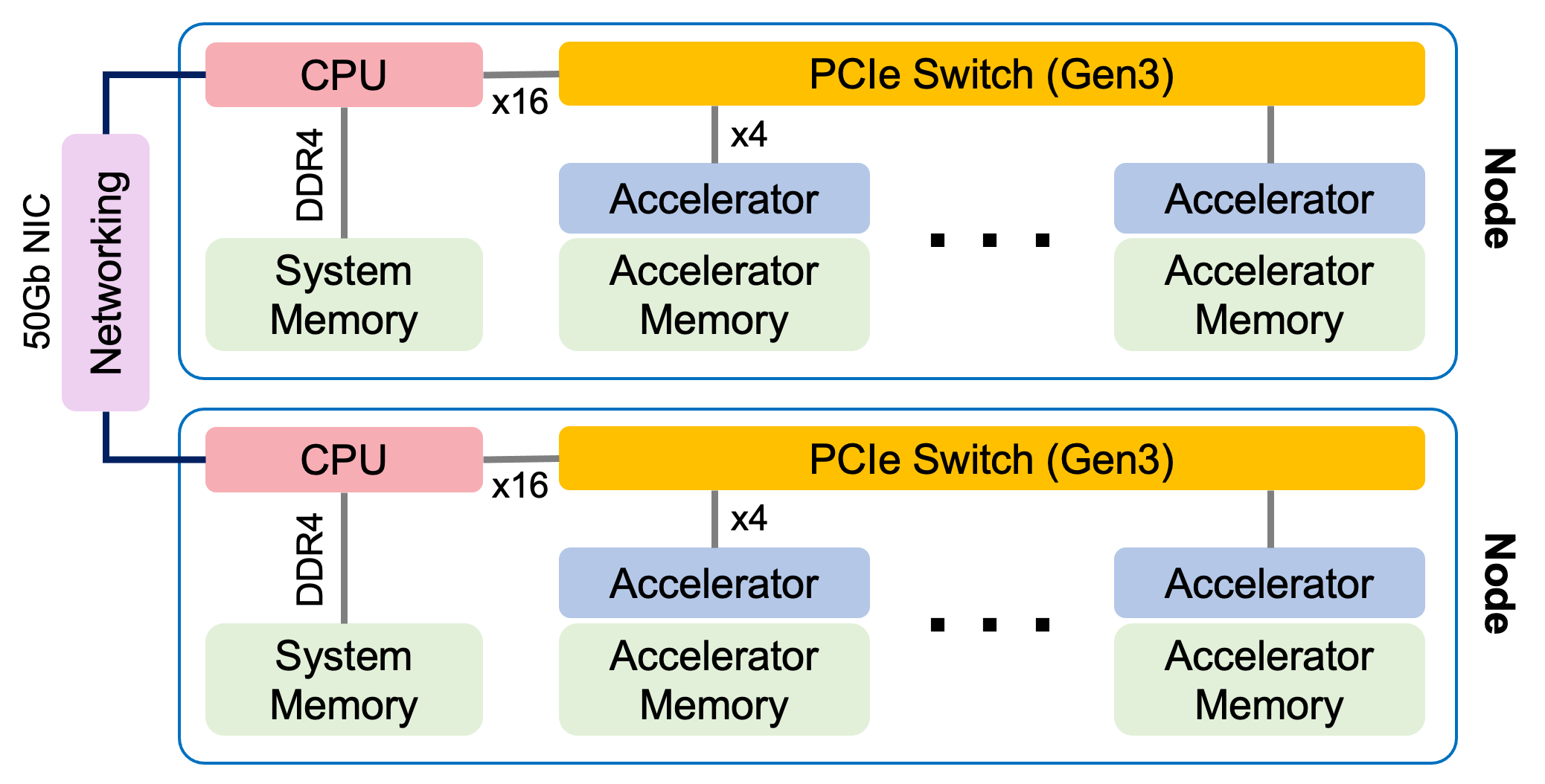}
    \vspace{-2.5mm}
    \caption{Updated Yosemite v2 Hardware System.}
    \vspace{-5mm}
    \label{fig:platform}
\end{figure}

We describe components of the system here. A set of six M.2 accelerators and a host CPU constitute a node. The host CPU is an Intel Xeon D with 64 GB RAM. Two sets of these nodes are installed into an updated Yosemite v2 \cite{yosemite_article} sled and are then connected to the TOR switch via a multi-host NIC. The system is designed as such to achieve high perf/W, which is our overall co-design goal. Each M.2 module has a power consumption of about 13W, and the power of the host CPU is amortized among six cards. 

In order to provide enough memory bandwidth for large recommendation and NLP models, we ensure that we can store these models in LPDDR memory. We opted for a multi-card system to provide enough aggregate card LPDDR memory capacity. Each of our cards provides 16 GB of LPDDR memory, and six of these cards, plus 64 GB on the host, provides about 160 GB, which we expect is enough to handle most of our models. Both host and accelerator memory can be used simultaneously when executing a model by splitting the net between host and accelerator. Given this multi-card setup, we could also distribute model weights among cards even for models that fit into a single card; the motivation here would be to fit into local memory on the card for efficiency. Furthermore, it gives us the option to scale to larger models in the future.

The cards are connected to each other via a PCIe switch, which is then connected to the host via x16 PCIe lanes. The PCIe switch consumes 13W. Each card has a x4 PCIe connection, and the PCIe switch enables for card-to-card communication without involving the host. That is, if one inference refers to a tensor which is produced by another inference, the data transfer is coordinated among the accelerator cards. This avoids the bottleneck on the x16 link to the host as well as host memory bandwidth if all cards communicate simultaneously. This provides enough bandwidth such that the PCIe bandwidth is not a bottleneck for most use cases. Further, to avoid potential NIC bottlenecks from data ingestion, we provisioned the system with an upgraded 50 Gbps bandwidth per node. 

\subsection{Accelerator Card}
\begin{figure}[t]
    \centering
    \includegraphics[width=0.3\textwidth]{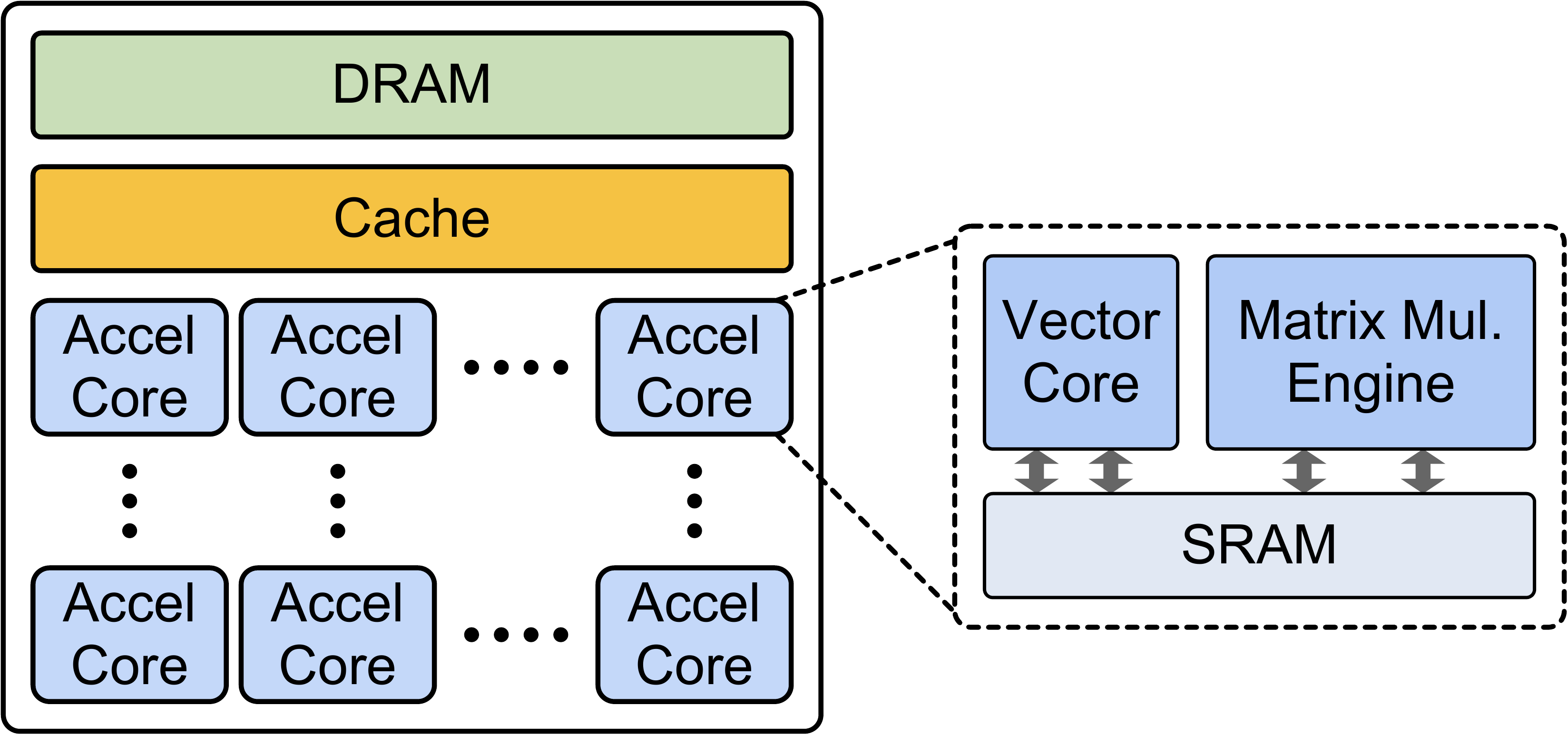}
    \vspace{-2.0mm}
    \caption{Accelerator Card.}
    \vspace{-5mm}
    \label{fig:nnpi}
\end{figure}

Most of our models spend a significant part of the time in dense compute layers, involving matrix multiplies and convolutions; a design aspect for us was to ensure that such operations could be efficiently computed. Furthermore, the weights for these compute layers in certain CV and recommendation system applications is in the range of tens of MB, so it is beneficial to have on-chip storage of this size. 

The Accelerator card (Figure \ref{fig:nnpi}) has a peak of 30-45 TOPS (int8) or 4-6 TFLOPS (FP16), depending on frequency. The card also has local SRAM memory inside of each Accel Core, and a shared cache, which may both be used to hold weights and activations of the network. The tight inter-connectivity of the system allows for a dense solution where we can also split model weights among cards to leverage the LPDDR and SRAM on all cards. 
\section{Software Design}
\label{section:software}

In order to realize the efficient execution of our target applications on the accelerator system, we designed a software system with a number of considerations in mind. First, we needed to ensure the software system was \textit{flexible} in that it is able to handle our current and future applications, support multiple frameworks including both Caffe2 and PyTorch, and be capable of targeting multiple types of backend hardware. Secondly, the system needed to be \textit{transparent} to our developers, which allows us to understand, debug, and test the system quickly and continuously. Third, the system needed to be \textit{programmable}, allowing us to configure and tune the specifics of the data movement and compute, and to add custom kernels when necessary. 

\begin{figure}[t]
    \centering
    \includegraphics[width=0.39\textwidth]{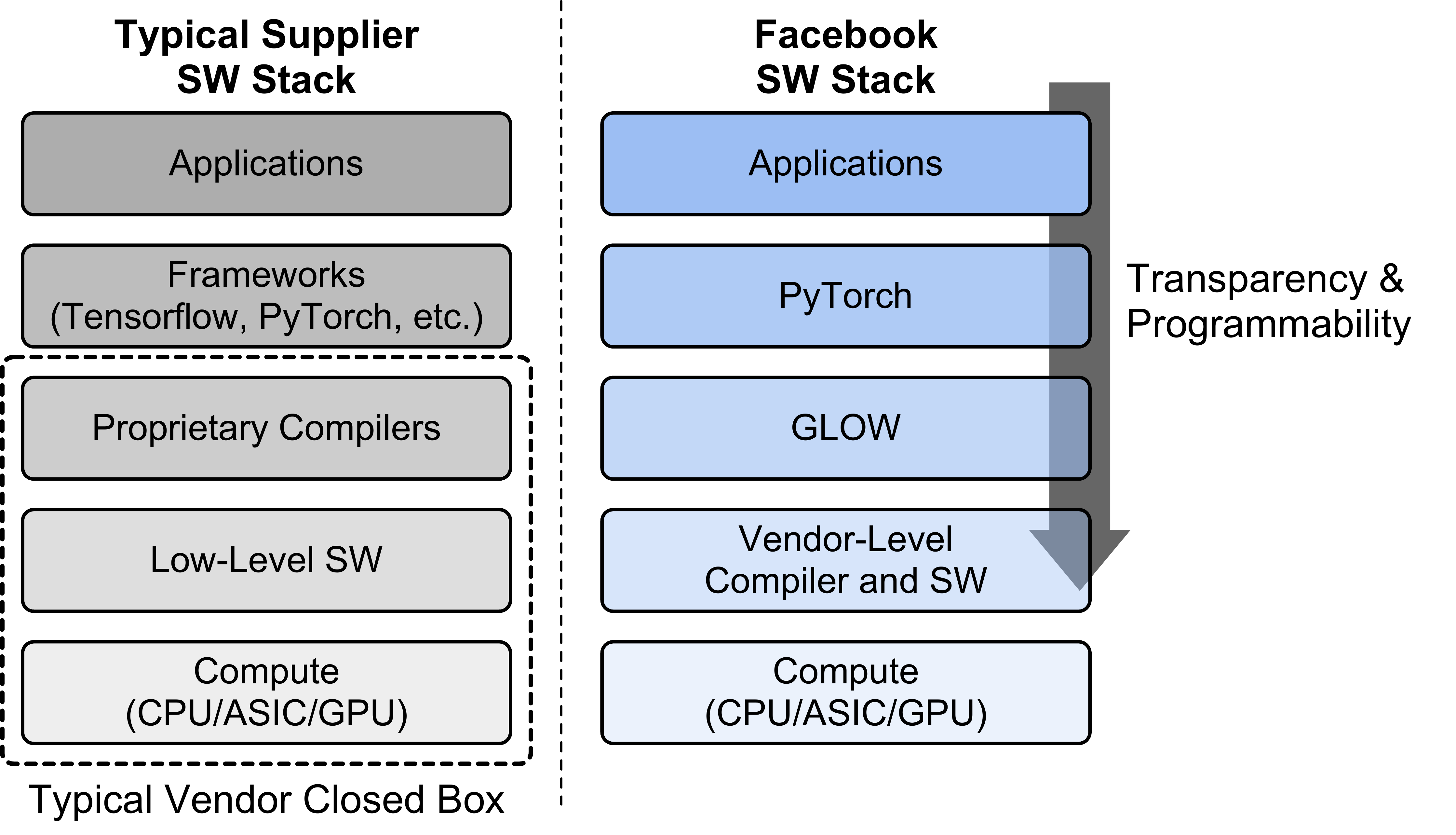}
    \vspace{-2mm}
    \caption{Software stack.}
    \vspace{-5mm}
    \label{fig:sw_stack}
\end{figure}

The left pane of Figure \ref{fig:sw_stack} shows the typical software stack proposed by AI ASIC vendors. In this stack, the customer provides applications through standard frameworks such as Tensorflow or PyTorch, and these are compiled and transformed via a vendor-provided set of tools to run on the hardware. While such systems provide enough flexibility to handle multiple applications and frameworks, they fail to provide the transparency and programmability desired for large-scale deployment. Furthermore, any optimizations which are dependent on a specific hardware backend are unable to be reused across multiple backends.

Instead we followed an approach shown in the right pane of Figure \ref{fig:sw_stack}, which improves the system in two respects. First, instead of compiling directly from the frameworks to the vendor compiler, we employed the Glow compiler \cite{rotem2019glow} to optimize the compilation graph before passing the intermediate representation to the vendor compiler. We also use the Glow runtime to manage resources and direct execution on the devices. Moving the device-agnostic portion of the compiler from black-box vendor software into a shared software stack improves transparency from our perspective. Second, for the portions of the systems remaining inside of the vendor’s stack, we worked closely with the vendor to expose these knobs in ways that we could leverage. Glow provides APIs to specify op placement and tensor placement. There are also special \textit{Custom Ops} which can be created in Glow, which can reference kernels written and compiled for the Accel Cores, allowing us to expand beyond the kernels provided. Glow supports many data types including float16 and bfloat16.

\subsection{Application Level}
Of the applications described in Section \ref{section:workloads}, the recommendation system models are currently implemented in Caffe2 and the Content Understanding models are implemented in PyTorch. At runtime, there is a custom binary which implements a service to respond to requests and execute inferences using the previously compiled network. 

\subsection{AI Framework Level}
Both Caffe2 and PyTorch have mechanisms for specializing certain portions of the computation graph for lowering to the accelerator through Glow. In the case of Caffe2, this procedure is called \textit{Onnxifi}, wherein the accelerated portion is compiled through the ONNX \cite{onnx} format into Glow, and executed as one large operator at runtime. For PyTorch we use the \textit{to\_backend} interface, which allows us to lower specific modules through Glow to the accelerator and replace these modules with logic to execute on the accelerator at runtime. In some cases such as CV detection models or the NLP application, certain operators are either not supported or perform poorly on the accelerator. In these cases we avoid lowering these operations to the accelerator and instead run on the host CPU.  

Because the accelerator and Glow both require shape information at compile time, we perform shape inference during this lowering stage. Additionally, to accommodate PyTorch models that use shape information as part of the model itself, for example for computing upsample output sizes, we treat these values as constant with respect to the compiled network and propagate any computations on them before compiling the final accelerator networks. In some cases where shapes vary at runtime, we use a combination of strategies, either switching between multiple compiled networks or setting an upper bound at compile time and padding input tensors where needed. 

\subsection{Glow Level} 

Once lowered from the Framework level to Glow, the computation graph is optimized using graph-level transformations. In the case of multiple-card execution, the graph may be partitioned at this level between cards both in model or data parallel fashion. Each device can be assigned to execute one or more partitions. Operators are also parallelized at this level in order to best fit the multiple Accel Cores on the chip. Numerous graph optimizations such as eliminating common sub-expressions or unnecessary conversions are also performed. After all of these optimizations, the graph is presented to the vendor software stack for code generation and device-specific optimization. We built support in Glow for passing hints to the vendor compiler which control operator-to-compute-unit mappings, tensor-memory mappings, and operator ordering. The operator-to-compute-unit mappings and operator ordering hints are set during compilation for recommendation system models. If the hints are not set, then the system falls back to default behavior of the vendor compiler.

The Glow runtime is able to manage multiple requests in a queue, and distribute them to multiple devices as the devices become available. The runtime is multi-threaded which allows for the handling and overlapping of multiple requests at a time, which is essential for high throughput. 

\subsection{Vendor Level}

The vendor compiler will translate the optimized compute graph for each partition and generate code specific to the low-level compute units inside the device. Decisions are made at this level pertaining to which of many optimized kernels to choose, which of several compute units to target, and whether or not to fuse or chain multiple ops at this level. The placement hints provided from Glow are followed where possible, and discarded if for any reason they are not satisfiable. Some examples of rejected hints: tensor-memory mapping hints may be rejected if the capacity of the specified memory is exceeded, and operator ordering hints may be rejected if they violate the network's data dependencies.
\section{Numerics}
\label{section:numerics}
To benefit from the high compute throughput on the accelerator and achieve real performance wins, we need to leverage low-precision numerics in the Matrix Engine or Vector Core, which provide orders of magnitude higher compute throughput than FP32. Low-precision model inference has been popular recently to mitigate the demand for  memory and compute resources. The challenge, however, is to bound the accuracy drop from the reduced precision. Low precision inference itself, especially for recommendation models, has a lot of interesting challenges that we don't have enough space for in this paper. We provide an overview here and more details will be described in a separate paper. Exploiting the low-precision numerics on the hardware is extremely challenging because we need to provide an accuracy guarantee for models from  different domains. Moreover, we also need to verify that the numerics behaviors of the low-precision model are as expected using the numerics support on the hardware. In this section, we first look at the application-level requirements and quantization workflow for the different models. And then we discuss the numerics validation for the low-precision models running on the accelerator. 

\subsection{Accuracy requirements}

\noindent\textbf{Recommendation systems}: The accuracy requirement for  recommendation systems is that the end-to-end online accuracy should be neutral using the low-precision models. The online accuracy metrics are service dependent and can only be measured in A/B testing or in production after serving a portion of user feedback. Therefore, an offline accuracy metric, which is easier to measure, is commonly used in early stages of evaluation. The offline metric is called normalized (cross) entropy (NE)~\cite{10.1145/2648584.2648589}. The maximum NE degradation tolerance is about 0.02\%-0.05\%, for low-precision recommendation models compared to the original FP32 model. In addition to the strict accuracy requirements, a challenge is that the recommendation models are updated periodically and frequently. Therefore, a robust quantization workflow without manual intervention is essential, and we cannot rely on int8 quantization working correctly for every case (hence the importance of reasonably high fp16 compute throughput as a fallback).

\vspace{1mm}
\noindent\textbf{Computer vision (image and video)}: There are a variety of models in the computer vision domain that exploit different model architectures as well as  accuracy metrics. Overall, the accuracy requirements for CV models are not as strict as for the recommendation models. For classification and detection models, commonly used accuracy metrics are top-1/top-5 scores or mAP (mean average precision) score where the accuracy threshold for quantization could be up to 1\%. For the backbone models that only produce embeddings to further feed downstream models, cosine similarity is used to evaluate the quality of the embeddings. In this case, we expect to achieve $\ge 98\%$ cosine similarity on the embeddings using low-precision models, but we should further check the end-to-end accuracies on the downstream models and enforce their requirements to be met. 

\noindent\textbf{Natural Language Processing (NLP)}: NLP workloads encompass a broad range of tasks, such as text embedding and classification, natural language understanding (NLU), and neural machine translation (NMT).  Text embedding models are backbone models that generate embeddings which are processed by downstream models.  Similar to CV applications, we use cosine similarity to evaluate the impact of low precision arithmetic, in addition to downstream model accuracy.  For classification tasks, e.g., NLU intent classification to identify a task and task parameters, top-1 accuracy is appropriate.  Finally, the BLEU (Bilingual evaluation understudy) score is commonly used to evaluate the translation accuracy for one pair of source and target languages. For a given NMT model, we need to check BLEU scores of all the most frequently used language pairs and make sure the accuracy drop is less than 0.1\%.  

\subsection{Quantization workflow}
Since quantization could incur accuracy loss for the model inference, we want to apply quantization only on the compute-heavy operators. So, the first step in the quantization workflow is to profile performance, which can identify the operators that are top bottlenecks in the end-to-end inference latency and are hence a priority for quantization. And then we explore different quantization schemes on the target operators, such as FP16, int8 or int4, etc. We search the quantization schemes with different precisions iteratively until the end-to-end accuracy can meet the application-level requirement. We use the per-layer quantization error as the feedback and try to increase the precision for those operators that could otherwise incur high quantization errors.  

For recommendation systems, the Fully Connected (FC) and Sparse Lengths Sum (SLS) operators account for most of the inference latency so we try to execute these operators in either the Vector Cores or Matrix Engines. The SLS operators are described briefly in Section \ref{section:workloads}. The embedding tables of recent recommendation models contain more than a hundred gigabytes of parameters. Using mixed precision of int8 and int4 for the embedding tables~\cite{guan2019posttraining} can greatly reduce the memory capacity and the model size without compromising the accuracy. The compute of SLS operators can also be done in FP16 in the Vector Cores without accuracy loss. For the FC operators, we try to use int8 quantization on as many operators as possible, but fall back to FP16 when the quantization error is too high. Usually we need to skip a few FC operators, including the last FC, in order to meet our requirement to be within the 0.05\% NE threshold.

For the CV and NLP models, int8 quantization usually works for most of the operators in the model, except the first convolution operator and the last non-linear operator. For the int8 operators, channelwise quantization is used for accuracy. And the rest of the model will run in FP16. The NLP results in this paper reflect FP16 execution.

\subsection{Numeric validation}

In addition to minimizing the accuracy loss from quantization, we need numerics validation to deploy the low-precision models on the accelerators. When the kernel implementations in the hardware are not fully transparent, we need numeric reference implementations to verify that the kernels are correct and functioning as expected. To achieve that, we independently implemented operators based on the vendor’s reference implementations. We performed comprehensive unit tests at the operator level as well as at net level in order to compare the numeric reference implementations against the accelerator. Full net tests are particularly important because they can expose behaviors only happening on groups of operators such as operator fusions, e.g. SLS + Layer Normalization, Dequantize + Swish + Quantize, etc.
Having the reference match bit-wise to the vendor allowed us to not have to think about error propagation, which in large networks can yield to meaningless error bounds for the final result. A nice property we have gotten out of this exercise is that the kernel implementations are deterministic and give the same result regardless of the kernel being chosen depending on the input shapes.

The numeric reference implementations can efficiently run on the CPU. After the numerics validation, the reference implementations are in sync with the accelerator so they can be used to run large scale evaluations on CPUs and reflect the device. This was useful for us to experiment with different int8 quantization schemes on full networks, for example. However, the numerics validation needs to be done for every software release from the vendor. Sometimes changes due to performance optimizations can also affect numerics. We need to guarantee that both accuracy and performance for the models of interest don’t degrade using the latest release.  

When bugs are found, we need to ask for a fix from the vendor and update the reference implementations accordingly. Therefore, in order to automate the numerics validation, we have open sourced the operator level unit tests~\cite{fakelowp} based on the reference implementations so that the vendor can do the tests independently for each new release. Furthermore, we have set up continuous accuracy monitoring for the production models with the numeric reference implementations to make sure the accuracy is consistent in all model versions during online training. 

\section{Performance Optimization}
\label{section:optimizations}

There is a significant amount of optimization effort involved in tailoring the system for high performance. This section will describe these optimizations, categorized into:

\begin{itemize}
    \item \textbf{Model level optimizations}: Changes that require collaboration with infra and modeling teams to make changes in the model, either modifying data types or modifying the compute graph,
    \item \textbf{Card-side optimizations}: Changes that require collaboration with compiler teams to make changes to the compute graph, or card-level improvements, and
    \item \textbf{System-level optimizations}: Optimizations encompassing the entire system, including host CPU, NIC and PCIe.
\end{itemize}

\subsection{Model level optimizations}
\noindent\textbf{Quantization and data-type changes}: One of the optimizations for recommendation models was int8 quantization, where we leverage the increased compute capabilities of the accelerator hardware for int8 dense compute operations over FP16, and also the reduced memory footprint and memory bandwidth demand. The details of this implementation are in Section \ref{section:numerics}. We also got benefits from converting the input dense features for recommendation models from FP32 to FP16, thereby halving the data transfers over the network and halving host memory bandwidth involved for these feature transfers.

\vspace{1mm}
\noindent\textbf{Network optimization with a better net split}: 
There are a number of considerations which influence the specific policy when splitting a net between the host CPU and the accelerator system. First, the split should ideally be chosen such that it reduces PCIe transfer time between host and accelerator. Second, the CPU system in this case is capable of executing small operations with lower latency than the accelerator system. Third, the portion of the net assigned to the accelerator should contain the bulk of the compute, and should be supported by the accelerator. Lastly, network bandwidth and latency need to be minimized, since network transfer time is a significant component of the end to end latency.

Recommendation model serving involves a number of inputs which are replicated across the batch dimension before further processing on the accelerator. It's important to balance the cost of additional broadcast ops on the accelerator, vs the alternative which is to execute the broadcast on the host and send redundant data over the PCIe link. We found it favorable to do these broadcasts on the accelerator. However, there are a number of broadcasts which are done, one for each embedding table, before the results are concatentated. These small individual broadcasts add up to a relatively high overhead on the card, so we instead concatenate these on the host CPU where latency is low, pass the non-broadcasted tensor over PCIe, and perform a single large broadcast on the card. 

Similar considerations influence our host-accelerator split in NLP and CV processing. For NLP, certain portions of the application are difficult to execute on the device, such as the portions which convert string input sequences from text to tensors consisting of indices into a vocabulary. However, since the bulk of the XLM-R compute is contained in 24 Transformer layers, we offload just these to the device and perform the rest of the network on the host CPU.  Additional optimizations enable the embedding step performed prior to the 24 Transformer layers on the accelerator as well. For CV detection models, we also keep operators on the host CPU which either are not supported yet on the accelerator or don't map fully to the vector or matrix portions of the card for example region proposal generation which includes non-maximum suppression. For video workloads, the video decode is run on the host CPU which can become the bottleneck and limit the overall throughput of the server. In such cases, we isolate the video decode on a separate server where we can potentially use specialized video transcoder ASICs to accelerate the video decode. 

\vspace{1mm}
\noindent\textbf{Variable-size data structure padding}:  As previously described, some accelerators require tensor sizes to be known \textit{a priori} at compile time, conflicting with the naturally variable length of input data such as text.  We resolve this by extending the string to tensor conversion with padding logic, by padding each input with pad tokens to a predefined maximum length.  To avoid creating excessive processing overhead associated with shorter inputs, we build multiple copies of the XLM-R model corresponding to multiple padding boundaries (e.g., 32, 64, 128 and 512 tokens).  This padding is performed as part of the CPU-resident portion of the network, enabling to pick the one version of the accelerator-resident portion corresponding to the padded data size, from among the several versions corresponding to different input sizes. 

\subsection{Card-side Optimizations}

\begin{figure}[t]
    \centering
    \includegraphics[width=1.0\columnwidth]{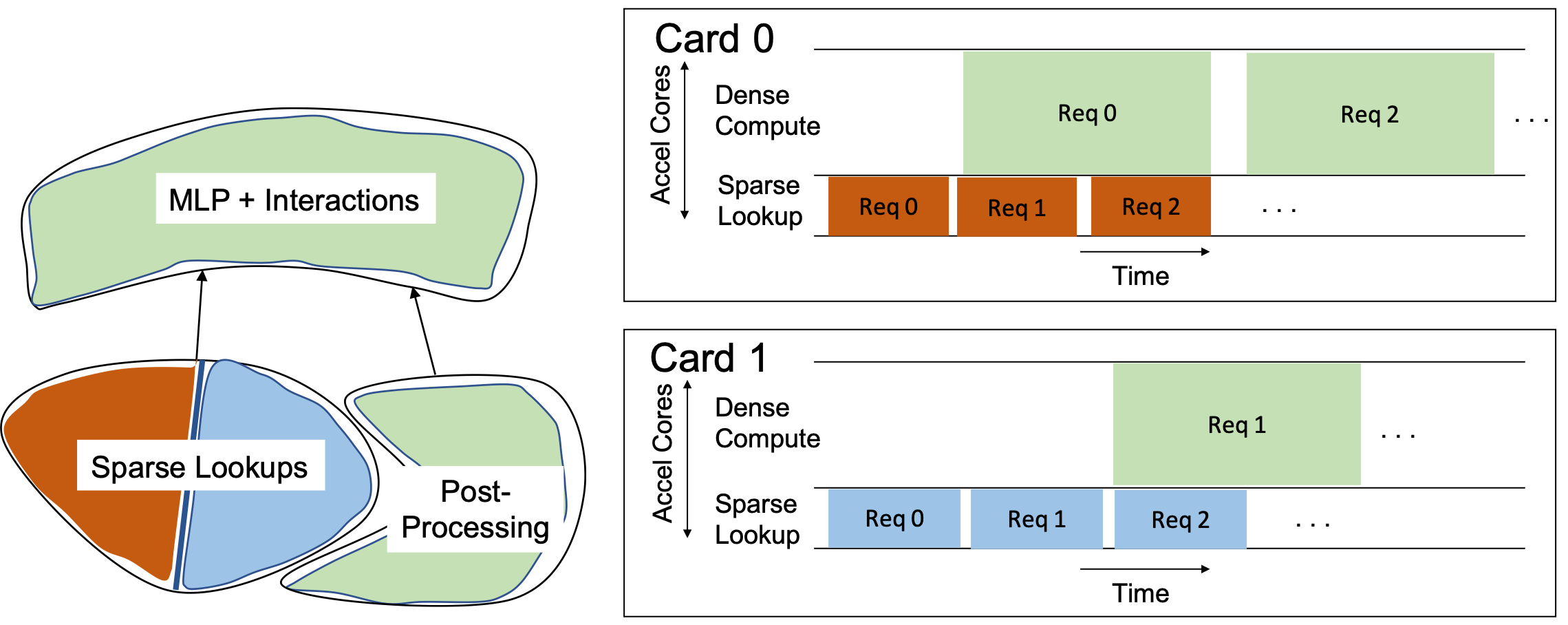}
    \vspace{-6mm}
\caption{Partitioning scheme for recommendation systems (left), and a demonstration of the pipelined execution of multiple requests of the partitioned net (right).}
    \vspace{-4mm}
    \label{fig:partitioning}
\end{figure}

\vspace{1mm}
\noindent\textbf{Model partitioning}: Recommendation models are too large to fit on the 16 GB of a single accelerator card and hence we partition the model across the six cards in a node. We developed a partitioning scheme specialized for recommendation system models, shown in Figure \ref{fig:partitioning}. In this scheme, we distribute the sparse embedding lookup tables across cards using model parallelism, and parallelize the dense compute across cards using data-level parallelism. A subset of Accel Cores is assigned to sparse lookups, and the rest to dense compute. For a given request, the sparse lookups from multiple cards must be communicated over PCIe to execute the dense portion of that request. Since multiple requests to the accelerator can be executed concurrently and pipelined, the steady state behavior of the system is to execute sparse lookups from one request at the same time as dense compute from another request. This allows us to scale to large recommendation system models without sacrificing throughput. 

Computer Vision and NLP models are so far smaller and fit on a single card. We have been using data parallelism for these models, which entails sending different images and sentences to different cards and executing them independently. One possible avenue for improvement in CV and NLP models would to be employ model partitioning over cards in order to potentially fit all model parameters into on-chip memory. 

\vspace{1mm}
\noindent\textbf{Resource allocation}: In the case of recommendation models partitioned using the scheme in Figure \ref{fig:partitioning}, there are multiple partitions running concurrently on a card. In this case, we must determine how many Accel Cores to allocate to each partition in order to balance their runtimes. Given that there are only a small number of possible splits, equal to the number of Accel Cores, we manually sweep the space of allocations and find generally using 1 in 3 cores for SLS to be a good balance. 

For the CV detection model, multiple networks are created when there are ops that are not currently supported on the card. This leads to the model being offloaded to the card via two nets, with unsupported ops running on the host CPU. Here we performed a similar manual search of core allocations across the two nets to achieve good card utilization.

\vspace{1mm}
\noindent\textbf{Parallelization and placement across Accel Cores}: If the compute graph doesn't have enough independent (i.e. parallel) operations available for execution on all the Accel Cores, we have to split operations in the graph to create the necessary parallelism in the graph. In order to do this, apply in Glow a heuristic policy of splitting ops according to the op type, dimensions, and predecessors, which works well for our networks. With this policy, the overall Accel Core utilization achieved is 78\% for the Non-SLS partition of recommendation system networks. In NLP, for example, we see a 2.6$\times$ speedup when parallelizing using this heuristic compared to not doing so. 

Once the nodes are parallelized, we also are able to schedule them explicitly on the Accel Cores using a placement hint API built into Glow, using list scheduling informed by a performance model learned by profiling. Performance improvements from explicit placement vary depending on the model, but generally does not exceed 10\%-20\% improvement for recommendation system models. For those models that are being optimized for throughput rather than latency, such as CV models, we can afford to use a smaller number of cores to perform a single inference and run multiple requests in parallel.

\vspace{1mm}
\noindent\textbf{Batching}: Batching is a common optimization that helps in three ways. First, large batches increase the amount of data level parallelism that can be used to distribute the computation across the cards. Second, batching increases data reuse of the weights tensors for the dense compute operations improving compute/bandwidth ratios for the operations. Finally, batching can amortize the cost of fixed overheads. Batches in inference can arise if there are independent items to evaluate in an inference request, or from combining multiple requests. Batching is complicated for NLP models because varying sentence lengths can lead to smaller sentences being batched with larger sentences, which leads to wasted compute. In the CV concept trunk model, increasing batch size from 1 to 4 increases performance by about 1.6-1.8X. Video inference does not use batching because the temporal aspect of the inference does not lend easily to batching. In our experience, a downside of larger batches for recommendation systems is that in addition to increasing latency, it reduces the total number of batches in flight which can reduce the performance of other portions of the system. Batch size must therefore be balanced with these other considerations to maximize to total system efficiency.

\vspace{1mm}
\noindent\textbf{Average pool optimization}: RegNet-Y CV models have global pooling as part of the architecture. If these are not optimized for all pooling sizes (including full image), we observed severe performance degradation. It is important therefore for CV models to have kernels which perform well for both large and small sizes. Optimizing average pool ops for a variety of pooling sizes reduced the runtime of these ops substantially and reduced the fraction of time spent in these ops from 44\% to 6\%. 

\vspace{1mm}
\noindent\textbf{Optimizing Sparse Lookups}: The execution latency of Sparse Lengths Sum (SLS) operators depends on the number of embedding table lookups, which is only known at runtime. See Section \ref{section:workloads} for a description of SLS in the context of recommendation systems. With our internal performance modeling framework, we can estimate the average number of embedding table lookups, annotate the information in the model, and use this information at compile time to help SLS operator load balancing across multiple cores within card and across multiple cards. With the length information, we reduced SLS partition latency by about 15\%-34\% for selected complex models and setups, compared to naive load balancing without the information. Furthermore, some SLS operators perform exactly one lookup on their embedding table by design. In this case, we can use a simple lookup kernel instead of a full SLS kernel to reduce the latency. 

\subsection{System level Optimizations}

\vspace{1mm}

\noindent\textbf{Partial tensor transfers}: In recommendation system models, each inference request has a different number of embedding lookup indices. However, many accelerators require all shapes to be static, and so input embedding lookups indices must be a priori set to some maximum size. Naively this means that unused indices will be sent. We added the capability to transfer only ``partial tensors'', i.e. only transferring those values in the tensor that are actually used, in order to significantly reduce PCIe traffic in the common case while still supporting the maximum number of indices.

\vspace{1mm}
\noindent\textbf{Command batching}: Recommendation system models come with many small independent inputs. To overcome overheads of performing many small transfers for each embedding table lookup, command batching can be used to combine many small transfers into a single large transfer.

\vspace{1mm}
\noindent\textbf{Removing host intermediary for transfers}: Using the recommendation system partitioning scheme described earlier, after the sparse lookups partitions are spread across many cards, all intermediate results must be co-located on the device performing the dense compute partition. Thus, either intermediates need to be transferred between devices, or the intermediate already exists on the device which performs the dense compute partition. Thus we added support for device-resident tensors and peer-to-peer tensor transfers, allowing the host to be removed from the communication of tensors between devices and reducing PCIe transfers by over half.

\section{Experiment Results}
\label{section:results}

\begin{table*}[t!]
\centering
\caption{Op breakdown.}
\vspace{-2mm}
\scriptsize
\renewcommand*{\arraystretch}{0.9}
\renewcommand*{\tabcolsep}{1.3pt}
\begin{tabular}{rrlrrlrrlrrlrrlrr}
\Cline{0.25ex}{1-2} \Cline{0.25ex}{4-5} \Cline{0.25ex}{7-8} \Cline{0.25ex}{10-11} \Cline{0.25ex}{13-14} \Cline{0.25ex}{16-17} \noalign{\vskip 1pt} 
\multicolumn{1}{l}{} & \multicolumn{1}{c}{\textbf{Recom.(5x)}} &  & \multicolumn{1}{l}{} & \multicolumn{1}{l}{\textbf{ResNeXt101}} &  & \multicolumn{1}{l}{} & \multicolumn{1}{l}{\textbf{FBNetV3}} &  & \multicolumn{1}{l}{} & \multicolumn{1}{l}{\textbf{RegNetY}}  &  & \multicolumn{1}{l}{} & \multicolumn{1}{l}{\textbf{ResNeXt3D}} &  & \multicolumn{1}{l}{} & \multicolumn{1}{l}{\textbf{XLM-R}} \\ \Cline{0.25ex}{1-2} \Cline{0.25ex}{4-5} \Cline{0.25ex}{7-8} \Cline{0.25ex}{10-11} \Cline{0.25ex}{13-14} \Cline{0.25ex}{16-17} \noalign{\vskip 1pt} 
FC & 30.9\% &  & \multirow{2}{*}{\begin{tabular}[c]{@{}r@{}}Channelwise-\\ QuantizedConv\end{tabular}} & \multirow{2}{*}{57.30\%} &  & \multirow{2}{*}{\begin{tabular}[c]{@{}r@{}}Channelwise-\\ QuantizedConv\end{tabular}} & \multirow{2}{*}{67.0\%} &  & \multirow{2}{*}{\begin{tabular}[c]{@{}r@{}}Channelwise-\\ QuantizedConv\end{tabular}} & \multirow{2}{*}{68.1\%} &  & Convolution3D & 18.4\%  &  & MatMul & 72.5\% \\
SLS & 27.0\% &  &  &  &  &  &  &  &  &  &  & MatMul & 13.3\% &  & Add & 3.0\% \\ \cline{4-4} \cline{7-7} \cline{10-10}
BatchMatMul & 8.8\% &  & Add & 37.40\% &  & Fused Conv\_Add & 27.2\% &  & Tile & 13.7\% &  & Convolution & 10.2\% &  & Concat & 2.1\% \\
Transpose & 4.3\% &  & ConvertTo & 2.50\% &  & ROIAlign & 2.7\% &  & AdaptiveAvgPool & 6.0\% &  & Add & 6.5\% &  & Transpose & 3.6\% \\
Quantize & 4.8\% &  & Quantize & 0.60\% &  & ConvertTo & 0.7\% &  & Add & 6.0\% &  & Transpose & 6.5\% &  & Gelu & 2.2\% \\
Dequantize & 3.6\% & & AdaptiveAvgPool & 0.20\% &  & Quantize & 0.5\% &  & Mul & 4.4\% &  & MaxPool & 6.1\% &  & Softmax & 3.3\% \\
Others & 20.6\% & \hspace{1.6mm} & Others & 2.0\% & \hspace{1.6mm} & Others & 1.9\% & \hspace{1.6mm} & Others & 1.8\% & \hspace{1.6mm} & Others & 39.0\% & \hspace{1.6mm} & Others & 12.4\% \\ \Cline{0.25ex}{1-2} \Cline{0.25ex}{4-5} \Cline{0.25ex}{7-8} \Cline{0.25ex}{10-11} \Cline{0.25ex}{13-14}  \Cline{0.25ex}{16-17}
\end{tabular}
\label{table:breakdown}
\vspace{-4mm}
\end{table*}

\begin{figure}[t]
    \centering
    \includegraphics[width=0.9\columnwidth]{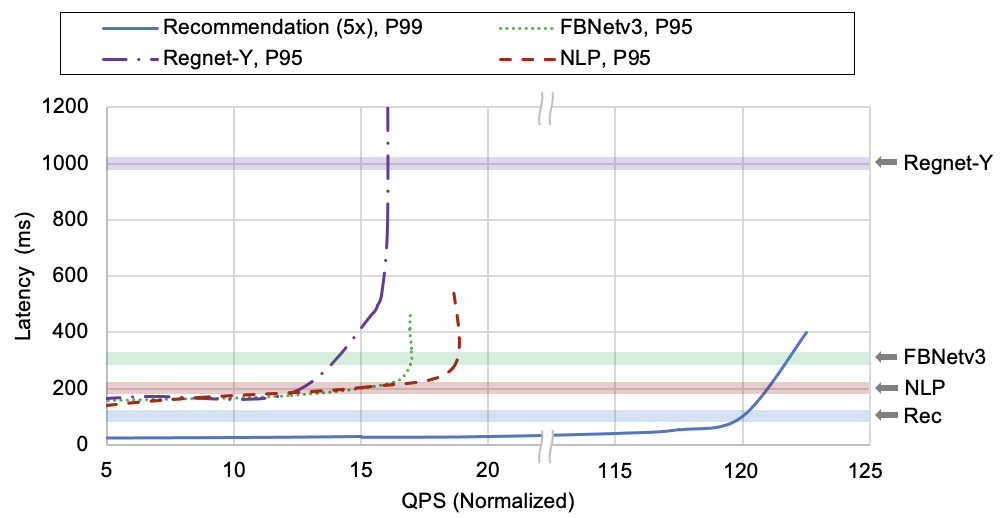}
    \vspace{-2.5mm}
    \caption{Latency and relative QPS of our models on the accelerator system. The bands represent latency requirements of the models.}
    \vspace{-5mm}
    \label{fig:latency_qps}
\end{figure}

In the following sections we showed how the increased compute and bandwidth of the co-designed accelerator can be used to run more complex and more accurate ML models which cannot be easily or efficiently run on CPUs. Figure \ref{fig:latency_qps} shows the latency and relative queries per second (QPS) throughput for the complex models which were introduced in Table \ref{table:models}, along with their corresponding latency constraints. The accelerator is able to serve all of these complex models within the latency budget. Note also that the recommendation system models are running at much lower latency and higher QPS per batch compared to the content understanding models. The op-level breakdown of runtimes for each model are shown in Table \ref{table:breakdown}. 

\vspace{1mm}
\noindent\textbf{Recommendation models}: For recommendation models, this corresponds to a 5$\times$ increase in GFLOPs compared to current models. The latency of these inferences ranges in the tens of milliseconds per batch. Given these strict latency requirements and the complex nature of this particular model, inference on this model is not easily or efficiently run on CPUs. The op-level breakdown in Table \ref{table:breakdown} shows a roughly equal split in runtime between Fully Connected and Sparse Lengths Sum, which are the largest components. 

\vspace{1mm}
\noindent\textbf{CV models}: For CV models where more time is spent in the dense convolution layers, we can expect higher efficiency. We are looking at two models for classification problems, a model based on the Resnext-101 trunk (7.8 B madds, or 15.6 GFlops/image) and a more complex model based on the Regnet-Y trunk (which is about 128 B madds, or 256 GFlops/image). For these models, we typically have an latency requirement of 300 ms - 1 sec, which are both met by the accelerator system.  We also run other types of CV models, including a CV detection model based on an efficient FBnetv3 based architecture. 

\vspace{1mm}
\noindent\textbf{NLP models}: We also focused on a complex text model based on a 24-layer XLM-R model variant. Here the focus is on maximizing throughput subject to latency constraints of around 200 ms. The accelerator system is able to meet these requirements at most QPS points. We should note that the NLP model currently runs in FP16; we anticipate int8 should yield about 1.6X (about 72\% of the time is spent in Matrix Multiplication layers, and these should speedup by $\sim$2X due to bandwidth considerations). Combining batching and int8 could help alleviate the bandwidth bound. Batching will also improve performance because it can increase arithmetic intensity beyond the low value shown in Table \ref{table:models}. One challenge is, in the case of varying sentence lengths naive batching approaches may combine smaller sentences with larger sentences, leading to wasted compute on zeros for the small sentences. In this case a smarter batching approach is favorable, which can combine sentences of similar lengths. 

\section{Inference Hardware Design Directions for future AI workloads}
\label{section:directions}

This section will discuss where we expect AI workloads are heading (both ranking and content understanding), the challenges that these workloads will pose, and what kinds of hardware features we'd hope to see in next generation inference hardware to address these challenges. 

\noindent\textbf{Much larger models}:
Our inference system is designed to support large embedding tables from Recommendation models in memory. 
However, both Recommendation and Content Understanding models (CV/NLP) will continue growing in size, requiring significant amount of memory and may not be able to fit in 96GB total of 6-accelerator card memory. To address this, we should consider hierarchical storage, where local LPDDR memory may be backed up by large-capacity persistent memory, e.g., Non-volatile Memory (NVM)\cite{eisenman2018bandana}. NVM is better for our serving needs than flash because of higher endurance ($>$60 pDWPD). Some of our models can be updated 10-20 times a day. A challenge is locality analysis for identifying candidate tables with large sizes and low bandwidth requirement. 

\noindent\textbf{More complex models}: 
For recommendation models, FLOPs will continue growing while the arithmetic intensity may not change much. Increasing batch sizes can improve the arithmetic intensity but it is challenging due to the constraint on serving latency. 
As models grow larger, performance is bounded by DRAM bandwidth because on-chip SRAM is not enough to fit both weights and activations. Alternatively, we can explore model parallelism (called FC sharding or FC pipeline parallelism) by splitting individual FC layers across subset of accelerators.
FC sharding avoids weight duplication to keep more weights (6x more with 6 cards) in SRAM, alleviating the bandwidth bottleneck. However, the mapping of such computation will become more complicated, which in return, requires complex but accelerator-friendly design of model architecture.
We should not also over-optimize accelerators for the specifics of current models but try to capture long term trends. For example, compared with our analysis a few years ago\cite{park2018deep}, the Transformer architecture is dominating language models instead of RNNs but both benefit from fast small-batch FC and large on-chip memory. RegNetY has replaced wide ResNeXt architecture in high-end CV models, but the trend such as deeper CNNs and use of blocks with group/depth-wise plus point-wise convolution has continued. The jury is still out there if the Transformer architecture will also dominate computer vision~\cite{dosovitskiy2020image}, but it can be a good idea to prepare for its implication of potentially unlocking even bigger CV models.

\noindent\textbf{Numerics support}:
Reduced-precision arithmetic is an important efficiency technique and has been a popular research topic particularly for edge devices. However, for models running on our data centers, we shouldn't expect every model and all of their compute intensive parts can run with low-precision integer operations, especially for the many recommendation models updated frequently with strict accuracy requirements.
Therefore, without decent FP16 compute throughput to fall back without too steep performance cliff, usability of accelerators can be greatly hampered.

We also found dynamic quantization\cite{pytorchdynamic} is an effective technique to improve accuracy and avoids the complexity of static quantization that needs profile activation tensor value distributions. Hardware support such as collecting min/max of output tensors can be useful technique to make dynamic quantization more efficient.  
Since recent training hardware supports bfloat16 \cite{tpu_cacm, kalamkar2019study, nvidiaampere}, bfloat16 support in inference accelerators can be useful to unify numerical format between training and inference.

\noindent\textbf{Importance of sparsity}: We have seen promising pruning results for recommendation\cite{ye2020adaptive} and CU models. We also believe, for huge language models like GPT-3, pruning can be a useful technique to keep the model size manageable. This calls for specialized HW support for sparsity.
It is easier to reduce memory (and on-chip memory) capacity and bandwidth with sparsity (e.g., pruned model is stored compressed and decompressed when loaded into local storage) than saving arithmetic operations\cite{zhang2020sparch}. The former is important sparsity support for recommendation and language models because we don't expect higher batch sizes (hence arithmetic intensity). 

\noindent\textbf{Importance of Content Understanding (CU) models}: Across many types of Content Understanding models, the number of model parameters and FLOPS as well as the size and quantity of inputs continue to grow exponentially. Accelerators are important to keeping up with this growth and ensuring that newer and more accurate models can continue to be used in production settings. Additionally combining multiple modalities such as images with text may also increase the complexity of models in the future. Many of these content understanding use cases contain relatively regular memory accesses and larger blocks of compute so they can benefit immensely from the accelerator's compute capabilities.

\noindent\textbf{Accelerator programmability}: As discussed throughout this paper, the domain of ML continues to rapidly evolve with the demand for novel operators quickly outpacing the development of a corresponding specialized compute engine. In order to ease the burden of deploying a new accelerator to production, we believe a shift towards accelerator designs with less heterogeneity in the types of computing units is necessary.  

\noindent\textbf{Embedding table compression}:
Compression is important to mitigate the growth of embedding table size, and this is another area where programmability is important.
For example, we didn't anticipate 4-bit quantization \cite{guan2019posttraining} and row-wise pruning when we were designing the accelerator system a few years ago ~\cite{park2018deep}.
Nevertheless, programmable vector cores in the accelerators were able to support these compression techniques.
In the future, clustering based approaches may further compress embedding tables, and HW support for fast look-ups to small tables can be helpful.

\noindent\textbf{Accelerated video decode}: 
Video decode constitutes a sizeable proportion of the video workloads. Having a specialized video decoder hardware in close proximity to the inference accelerator would help reduce the overall latency and the network/PCIe/memory bandwidth. The video decoder could be integrated with the inference accelerator or it could be a separate device on the PCIe. Such a decoder would need the ability to output video clips at multiple resolutions and should be programmable so that the decoded frames can be processed with techniques such as normalization, stabilization and optical flow computation.

\noindent\textbf{Processing-in-memory}:  We've investigated applying processing-in-memory (PIM) to our workloads and determined there are several challenges to using these approaches. Perhaps the biggest challenge of PIM is its programmability. It is hard to anticipate future model compression methods, so programmability is required to adapt to these. PIM must also support flexible parallelization since it is hard to predict how much each dimension (the number of tables, hash size, or embedding dimension) will scale in the future (while for example, TensorDIMM only exploits parallelism across embedding dimension). In addition, TensorDIMM~\cite{kwon2019tensordimm} and RecNMP~\cite{ke2020recnmp} use multi-rank parallelism within each DIMM but typically the number of ranks per DIMM is not enough to provide significantly high speedup.

\section{Related Work}
\label{section:relatedwork}

ML models are good hardware acceleration candidates as they are 
deterministic as well as very demanding in terms of compute and memory bandwidth. 
Accordingly, major institutions such as 
Google \cite{tpu_v1, tpu_cacm}, 
Microsoft \cite{fowers2018a, chung2018serving}, 
Amazon \cite{inferentia}, and NVIDIA \cite{nvidiat4} 
have successfully developed and deployed ASIC, FPGA, or GPU-based inference systems to effectively accelerate their ever growing production ML models including MLPs, CNNs, RNNs, and LSTMs. 
From academia, various ML accelerators targeting 
compute-heavy CNN/DNNs \cite{isscc_2016_chen_eyeriss,chen2014dadiannao,eie}, 
MANNs \cite{mnnfast, 9269453, manna},
recommendation models \cite{hwang2020centaur}, 
an arbitrary DNN using a reconfigurable hardware or a systematic methodology
\cite{kwon2018maeri,7551399, sharma2016high, kwon2020heterogeneous}, 
and generic ISAs \cite{liu2016cambricon} have been proposed. 
In addition, techniques to optimize ML graph \cite{jia2019taso},
reduce the end-to-end latency via scheduling \cite{gupta2020deeprecsys},
improve the accelerator utilization with multi-tenancy \cite{baek2020multi,9065590}, 
scale up beyond a single system \cite{9218528} with better partitioning \cite{song2020accpar},
and better handle a particular component of an ML model such as embedding lookups 
\cite{eisenman2018bandana,ke2020recnmp,kwon2019tensordimm}
have been discussed.

In addition, the characteristics of various ML models have been discussed extensively
\cite{park2018deep, naumov2019deep, instagram_model, arch_implications_facebook_dnn_rec,lui2020understanding, hsia2020crossstack} to better understand their nature and design highly effective systems.
Collaborative benchmarking efforts such as MLPerf \cite{9001257} are making standardized ML models more accessible.

While these studies provide great insights on how to design and utilize an accelerator system, they are often limited to discussing only the microarchitectural details without the full view of the system including the host, and/or focusing on a specific type/part of the ML model architecture. The value of this work is in sharing a holistic HW/SW co-design experience of developing and deploying an accelerator system: how to build a system system that can effectively handle a diverse set of models (e.g., CV, NLP, recommendation), and how to best tailor the models to fully leverage the given system.

\section{Conclusion}
\label{section:conclusion}

This paper provided an overview of Facebook's hardware/software inference accelerator ecosystem and accelerator deployment. In order to meet increasing demands for inference compute in our data centers, we co-designed an accelerator system with our specific recommendation system, CV, and NLP workloads in mind. The accelerator system contains six accelerator cards connected over PCIe to form a system capable of delivering 180-270 TOPS in a 91W power envelope, including the PCIe switch. The system is based on an open hardware platform and open source software system. After significant software optimizations, the accelerator system is able to process significantly larger models for our targeted workloads more easily and efficiently than CPUs.


\bibliographystyle{IEEEtranS}
\bibliography{refs}

\end{document}